\documentstyle[prl,aps,epsf,psfig]{revtex}
\newcommand \beq{\begin{eqnarray}}
\newcommand \eeq{\end{eqnarray}}
\newcommand{\set}[2]{\newcommand{#1}{#2}}
\set{\pa}{\partial \over \partial\, }
\set{\ba}{\bar }
\set{\e}{\epsilon }
\set{\intp}{\int{dp\over (2 \pi)^3}}
\set{\ppl}{(p+\frac q 2)}
\set{\pmi}{(p-\frac q 2)}
\set{\leftvector}{\stackrel{\leftarrow}{\partial }}
\set{\rightvector}{\stackrel{\rightarrow}{\partial }}
\begin{document}
\twocolumn[\hsize\textwidth\columnwidth\hsize
           \csname @twocolumnfalse\endcsname
\title{New collective mode due to collisional coupling}
\author{K. Morawetz $^{1,2}$, R. Walke $^{1,2}$, U. Fuhrmann $^1$}
\address{$^1$
Fachbereich Physik, Universit\"at Rostock,
18051 Rostock, Germany\\
$^2$ Laboratorio Nazionale Del Sud, Via S. Sofia 44,95123
Catania, Italy
}
\date{\today}

\maketitle
\begin{abstract}
Starting from a nonmarkovian conserving relaxation time approximation for
collisions we derive coupled dispersion relations for asymmetric
nuclear matter. The isovector and isoscalar
modes are coupled due to asymmetric nuclear meanfield acting on
neutrons and protons differently. A further coupling is observed
by collisional correlations. The latter one leads to
the appearance of a new soft mode besides isoscalar and isovector modes in the system. We
suggest that this mode might be observable in asymmetric
systems. This soft mode approaches the isovector mode for high temperatures.
At the same time the isovector mode remains finite and approaches a constant value at higher temperatures showing a transition from zero sound like damping to first sound. The damping of the new soft mode is first sound like at all temperatures.
\end{abstract}
\pacs{21.60.Ev, 24.30.Cz, 24.60.Ky, 82.20.Mj}
\vskip2pc]

The investigation of collective excitations in asym\-metric nuclear
matter is of current interest for experiments with nuclei far from $\beta$- stability, \cite{CTL97} and citations therein. 
We consider here a Fermi gas
model consisting of a number
of different species (neutrons, protons, etc) interacting with
the own specie and with other ones. The interaction between different
sorts of particles is important to
consider if we want to include friction between different
streams of isospin components. Especially the isospin current may
not be conserved by this way. We neglect explicitly shell effects and concentrate only on bulk matter properties. Let us start with a set of coupled
quantum kinetic equations for the reduced density operator
$\rho_a$ for the specie $a$
\beq
\partial_t \rho_a(t) =i[\rho_a,{\cal E}_a+{\cal
U}(t)_a]-\sum\limits_b\int\limits_0^t
dt'{\rho_a(t')-\tilde \rho_b(t') \over \tau_{ab}(t-t')}\label{kin}
\eeq
where ${\cal E}={\cal P}^2/2m$ denotes the kinetic energy
operator and ${\cal U}$ the mean field operator and the external
field which is assumed
to be a nonlinear function of the density. We have approximated the
collision integral by a non-Markovian relaxation time \cite{FMW97}.
The memory effects condensed in the frequency dependent relaxation time
turned out to be necessary to reproduce damping of zero sound
\cite{AB92,MTM96}. It accounts for the fact that during a two particle
collision a collective mode can couple to the scattering process.
Consequently, the dynamical
relaxation time represents the physical content of a hidden three
particle process and is equivalent to the memory effects. 

We have further assumed
the relaxation with respect to the local equilibrium $\tilde \rho_b$ of any specie in the system. The relaxation of the actual distribution of specie $a$ is driven by the local equilibrium of all the other components. The cross coupling is realized by nondiagonal relaxation times $\tau_{ab}$.
We specify the local equilibrium 
by a small deviation of the chemical potential of specie $a$ \cite{M70} compared with the global equilibrium
\beq
<k |\tilde\rho_a|k'>=f_a(k)\delta_{kk'}-{f_a(k)-f_a(k') \over
\e_a(k)-\e_a(k')} \delta \mu_a(k-k')
\eeq
with $<k|{\cal E}|k'>=\e(k)$. The equilibrium distributions
$<k|\rho_a^0|k'>=f_a(k)\delta_{kk'}$
are the corresponding Fermi functions with chemical potential
$\mu_a$ and the normalization to density $n_a=2 \intp f_a(p)$. The local equilibrium specified by $\delta \mu_a$ is
determined if we impose the density balance to be fulfilled separately
for each specie current $J_a$ which reads in Wigner coordinates
$k=p+q/2$ and $k'=p-q/2$
\beq
\omega \delta n_a(q,\omega)=q \delta J_a(q,\omega).\label{con}
\eeq
From this equation we derive the following matrix equation for $\delta \mu_a$
\beq
\Pi_a(q,0) \delta \mu_a=\sum\limits_b \left \{1 \over
\tau\right \}^{-1}_{ab} \left ( B_b \delta(q)+{\delta n_b \over
\tau_b}\right )\label{mu}
\eeq
where $B_a=\sum\limits_b
\frac{n_b-n_a}{\tau_{ab}}$,
\beq
\frac{1}{\tau_a}&=&\sum\limits_b\frac{1}{\tau_{ab}},
\label{t1}
\eeq
and the partial polarization function of specie $a$ is
\beq
\Pi_a(q,\omega)=2 \intp\frac{f_a\ppl-f_a\pmi}{\e_a\ppl-\e_a\pmi-\omega}.
\eeq
The factor 2 in front of the integral accounts for the spin
degeneracy. Eq. (\ref{mu}) generalizes eq. 6 of \cite{M70} to multicomponent 
systems.

Now we linearize the kinetic equation (\ref{kin}) around the total 
equilibrium $\rho_a^0$ which we express as
\beq
\delta \rho_a=\rho_a-\rho_a^0=\rho_a-\tilde \rho_b+\tilde
\rho_b-\rho_a^0.
\eeq
This
provides us with the density variation $\delta n$ due to an
external field $U^{\rm ext}$ which
is connected with the polarization of the system via
$\Pi \delta n=U^{\rm ext}$. The obtained polarization function has now
a matrix structure. The poles of the polarization function represent
the collective modes in the system.
For a two component system, e.g. neutrons with density $n_n$ and
protons with density $n_p$, we have the density variation of the
mean field 
\beq
\delta U_a=\alpha_{an} \delta n_n+\alpha_{ap} \delta n_p
\eeq
and we obtain for the collective modes the dispersion relation
\beq
&&(1-\Pi_n^{\rm M} \alpha_{nn})(1-\Pi_p^{\rm M}
\alpha_{pp})\nonumber\\
&&-(D_{np}+\Pi_n^{\rm M} \alpha_{np})(D_{pn}+\Pi_p^{\rm
M} \alpha_{pn})=0\label{dism}.
\eeq
The generalization of the Mermin polarization function \cite{M70} to
a multicomponent system is derived with the inclusion of nonmarkovian (frequency dependent) relaxation times as
\beq
\Pi_a^{\rm M}&=&{\Pi_a(\omega+{i\over \tau_a})\over
1-{i \over \omega\tau_a +i} (1- C_{aa})}.
\eeq
An additional coupling in (\ref{dism}) occurs due to asymmetry and
collisions
\beq
D_{np}={\tau_n\over \tau_p}{C_{np}\over C_{nn}-i \omega \tau_n}.
\eeq
The $D_{pn}$ are given by interchanging sort indices.
The matrix $C_{ab}$ is expressed as
\beq
C_{ab}=\sum\limits_c \left\{ 1\over\tau \right \}_{ac}
{\Pi_c((\omega+{i\over \tau_a}){m_a\over m_c})\over \Pi_c(0)}
\left\{ \frac{1}{\tau}\right\}^{-1}_{cb}.\label{c1}
\eeq
The term $D_{np}$ is vanishing for
symmetric matter, i.e. for equal densities of species as well as
for vanishing collision integral $\tau\rightarrow \infty$. Therefore
we call this term asymmetry coupling term further on.

The
dispersion relation (\ref{dism}) is similar to the one derived
recently in \cite{CTL97} if we neglect the collisional coupling
$D_{np}$. The latter one has also been discussed in plasma two- stream 
instabilities \cite{HW91}. Here we present a more general dispersion including correlational coupling.
The polarization function includes collisions within a conserving approximation
\cite{HPR93}.

Before we apply this dispersion relation to nuclear matter let us consider a special case. We assume symmetric nuclear mean fields
$\alpha_{nn}=\alpha_{pp}=\alpha_1$ and
$\alpha_{np}=\alpha_{pn}=\alpha_2$ and neglecting the
collisions we obtain
\beq
1-(\alpha_{1}\pm\alpha_{2}) {\Pi(\omega+{i\over \tau})\over
1-{i \over \omega\tau +i} (1- {\Pi(\omega+{i\over
\tau}) \over \Pi(0)} )}=0.\label{dis4}
\eeq
These are the decoupled dispersion relations
for the isovector mode $\alpha_{1}-\alpha_{2}$ and the
isoscalar mode $\alpha_{1}+\alpha_{2}$.

Summarizing, we see that there appear two kinds of coupling: (i) The
coupling of modes between isovector and isoscalar ones due to different
mean fields and (ii) an explicit correlational coupling of
asymmetric nuclear matter due to collisional correlations, which
is condensed in $D_{np}$.
We have presented a general dispersion relation for the
multicomponent system including known special cases. 

In the
following we will apply this expression for the damping of giant
dipole resonances in asymmetric nuclear
matter.
We assume a wave vector for giant dipole resonances
corresponding to
the Steinwedel and Jensen \cite{SJ68} model $q={2.1 \over R}$ with the nuclear radius
$R=1.13\, A^{1/3}$ fm connected to mass number $A$. This wave vector
has very low values compared with the Fermi wave
vector. Therefore it allows us to expand the Mermin polarization function
with respect to small $q v_c/\omega$ ratios where $v_c$ is the sound
velocity. The frequency dependence of the dynamical (memory) relaxation
times is derived using a Sommerfeld expansion \cite{FMW97}
\beq
{1\over \tau_{ab}(\omega)}&=&{1\over \tau_{ab}(0)} \left (1+\frac 3 4
\left ({\omega \over \pi T}\right )^2 \right )\label{mem}
\eeq
for $a,b$ neutrons or protons respectively. The markovian relaxation time was given in terms of the cross section $\sigma_{ab}$ between specie $a$ and $b$ as $\tau^{-1}_{ab}={4 m\over 3 \hbar^3}\sigma_{ab} T^2$.

The dispersion relation
(\ref{dism}) takes then the form of a polynomial of tenth (six) order 
corresponding to the inclusion of memory (in)dependent relaxation times via
(\ref{mem})
\beq
&&0=(\omega ( \omega +{i \over\tau_n})-c_{nn}^2 q^2)
(\omega ( \omega +{i \over\tau_p})-c_{pp}^2 q^2)\nonumber\\
&&
-\left(c_{np}^2+i{\ba c_{np}^2 \over ( \omega +{i \over\tau_n})
\tau_p}\right )
\left (c_{pn}^2+i{\ba c_{pn}^2 \over ( \omega +{i \over\tau_p}) 
\tau_n}\right )
q^4\nonumber\\&&\label{order}
\eeq
with the definition (\ref{t1}) and (\ref{mem}).
Here the partial sound velocities are
\beq
c_{ab}^2&=&\alpha_{ab} {n_a(\mu_a)\over m}\nonumber\\
\ba c_{ab}^2&=&{1 \over m} {E_n-E_p \over {\tau_{np}\over
\tau_{pp}}-{\tau_{nn}\over \tau_{pn}}}\nonumber\\
\label{disend}
\eeq
and denoting the Fermi energy with $\epsilon_f^a$
\beq
E_a&=&T {f_{3/2}({\rm e}^{\beta \epsilon_f^a})\over f_{1/2}({\rm e}^{\beta \epsilon_f^a})}.
\eeq
We observe that the new collisional coupling represented by $\bar c_{ab}^2$ vanishes if either the diagonal friction $\tau^{-1}_{aa}$ or the nondiagonal friction $\tau^{-1}_{ab}$ vanishes or the system is symmetric $E_n=E_p$. This underlines that we have a new coupling due to collisional correlations and asymmetry.

\begin{figure}
\centerline{\psfig{file=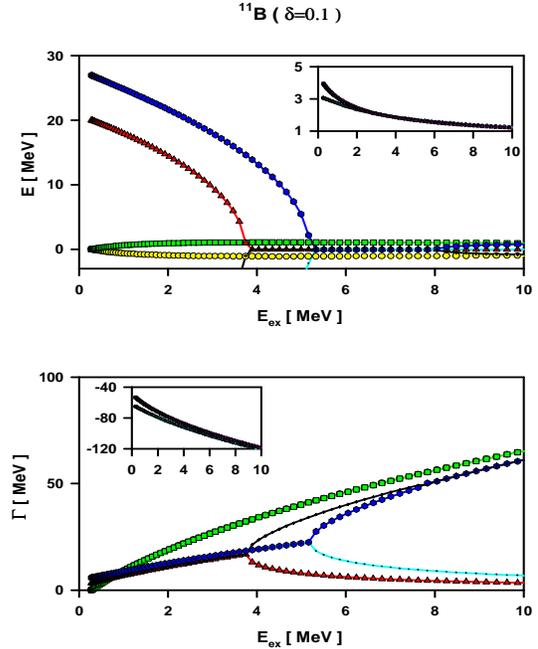,height=10cm,width=8cm,angle=0}}
 \caption{The centroid energy (above) and the damping width
 (below) of collective modes vs. excitation energy $E_{\rm ex}=\frac 1 4 (E/A)$ (above Fermi energy) 
found for $^{11}B$ situation.
 The isovector mode (rhombus) vanishes at about $5$ MeV while
 the corresponding isoscalar mode (triangles) already disappears at
 $4$ MeV. Connected with the disappearing the corresponding damping becomes 
twofolded. This is explained by the symmetric negative modes plotted as thin lines which disappear as well. A soft
 mode (squares) arises and is connected with an continuous increasing
 damping width. The inset gives the instable modes separately. \label{1} }
\end{figure}

We use as an illustrative example the following mean field
parameterization of Vautherin \cite{VB72,BV94}
\beq
U_a&=&t_0 \left [ (1+{x_0\over 2}) (n_n+n_p)-(x_0+\frac 1 2) n_a \right ]
\nonumber\\
&&+{t_3 \over 4} ((n_n+n_p)^2-n_a^2)
\eeq
with $n_a=n_n,n_p$ the density of neutrons or protons, respectively.
The corresponding mean field deviations can easily be computed via 
$\alpha_{ab}=\partial U_a/\partial n_b$. The Coulomb interaction 
leads to an additional contribution for the proton meanfield
\beq
U_p^{C}(q)={4 \pi e^2\over q^2} n_p(q).
\eeq
The here used model parameters reproduce the Weizs\"acker formula
\beq
{E\over A}= -a_1 +{a_2\over A^{1/3}}+{a_3 Z^2\over A^{4/3}} + a_4 \delta^2
\eeq
by the volume energy $a_1=15.68$ MeV, Coulomb energy $a_3=0.717$ MeV and the symmetry energy $a_4=28.1$ MeV with the asymmetry parameter
\beq
\delta={n_n-n_p\over n_n+n_p}.
\eeq
In figure \ref{1} we have plotted the solution of the dispersion
relation (\ref{order}) for $^{11}Be$ versus excitation energy. The kinetic energy is
linked to a temperature within the Fermi liquid model via
Sommerfeld expansion 
\beq
{E\over A} &=&\frac 3 5 \epsilon_f \left (
{(1+\delta)^{5/3}+(1-\delta)^{5/3}\over 2} \right .\nonumber\\
&+&\left . {5\over 12}\pi^2 ({T\over T_f})^2 {(1+\delta)^{1/3}+(1-\delta)^{1/3}\over 2} \right ).\label{fer}
\eeq
This connection between temperature and excitation energy is only valid for a continuous Fermi liquid model. For the small nuclei like $^{11}Be$, the concept of temperature is questionable. Some improvement one can obtain by the definition of temperature via the logarithmic derivative of the density of states \cite{BM69}
\beq
T^{-1}=\frac{1}{\rho}{\partial \rho\over \partial E_{\rm ex}}=-\frac  5 4 E_{\rm ex}^{-1}+\pi ({A\over 4 \epsilon_f E_{\rm ex}})^{1/2}
\eeq
which provides $E_{\rm ex}\approx \frac 1 4 (E/A)$ in comparison with (\ref{fer}) for small temperatures. We use this temperature to demonstrate possible collective bulk features in an exploratory sense. Of course, the surface energy and shell effects cannot be neglected for realistic calculations.

In figure \ref{1} we plot the occurring modes for a weak asymmetric case of 
$^{11}B$.
We observe that the isovector mode is decreasing with increasing
excitation energy and vanishes at about $5$ MeV. The corresponding
isoscalar mode is already vanishing at $4$ MeV. In general all energies 
appear as symmetric solutions with positive and negative energy where negative energy solutions are ruled out as unphysical. The corresponding damping is degenerated up to the point of vanishing energy. Above this temperature the damping of isoscalar and isovector modes become twofolded.

Besides the standard isovector and isoscalar modes
we observe a
build up of a very soft mode with a centroid energy around $1$ MeV.
This mode appears due to the collisional coupling $\bar c_{ab}^2$ of (\ref{disend}).
When we turn off the relaxation times, i.e. the collision integral,
this mode is vanishing as well as in symmetric nuclear matter, see discussion after (\ref{disend}). It shows that this mode appears due to
collisional coupling of isovector and isoscalar modes. The
corresponding damping of the crossed mode is continuously increasing with temperature.

Due to the consideration of memory (frequency dependent) relaxation times we observe further an unstable mode shown in the inset of figure \ref{1}. If we neglect this memory effect we would not observe this instable modes but the soft mode described above remains.

\begin{figure}
\centerline{\psfig{file=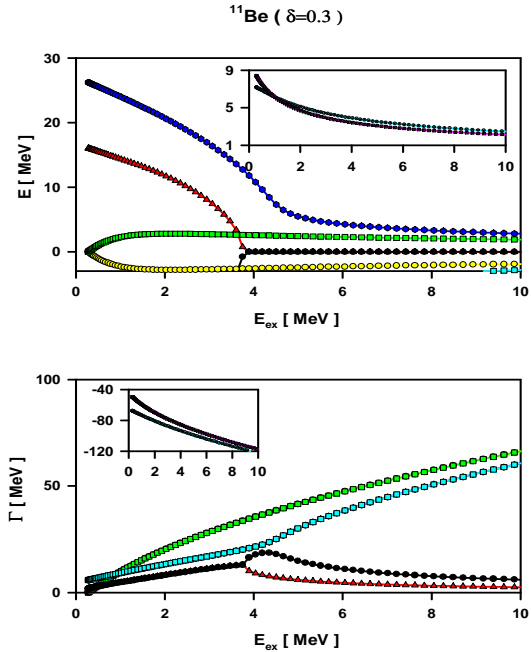,height=10cm,width=8cm,angle=0}}
 \caption{The centroid energy (above) and the damping width
 (below) of collective modes vs. excitation energy (above Fermi energy) 
found for $^{11}Be$ situation with the same set up as figure \protect\ref{1}.
The isovector mode (rhombus) decreases and remain constant at about $5$ MeV while
 the corresponding isoscalar mode (triangles) already disappears at
 $4$ MeV. A soft
 mode (squares) appears with higher centroid energy than in $^{11}B$ of figure \protect\ref{1} while the
 damping width is the same. \label{2} }
\end{figure}

Looking at a more asymmetric situation of $^{11}Be$ in figure \ref{2} we see that the centroid energy of the third mode becomes larger while the damping remains the same. A new feature appears for the isovector mode. While in weak asymmetric systems the mode disappear sharply at a certain temperature, we see now a decrease and a convergence towards the new soft mode. This suggest that the new soft mode should be of isovector character. This behavior of the isovector mode is connected with no transition to a twofolded damping we have seen for the more symmetric case. Instead the damping of isovector modes show a transition from a $T^2$ temperature behavior typical for zero sound damping towards a $E^{1/4}$ behavior typical for first sound damping. We like to pronounce that in symmetric nuclei we did not observe this transition because the mode itself, i.e. the centroid energy is vanishing at the point where this transition behavior of the damping occurs. In asymmetric nuclear matter it should be possible to observe this transition from zero to first sound because the energy of the isovector mode remains finite. The twofold damping of isoscalar mode is modified with respect to the symmetric case but not removed.

One may argue whether this third mode
can really appear in the system. A simple consideration may
convince us about the possible existence of such mode. Let us
assume a coupled set of two type of harmonic oscillators (neutrons
and protons) interacting between the same sort of particles with strength $k_n$
and $k_p$, respectively and between different sorts with $k_{np}$.
Let us choose for simplicity only two neutrons and two protons.
Then
we obtain the coupled system of harmonic oscillators with
frequencies $\omega_n^2=k_n/m$, $\omega_p^2=k_p/m$ and
$\omega_{np}^2=k_{np}/m$. The solution yields
three basic modes in the system, i.e.
$\omega^2=2(\omega_n^2+\omega_{np}^2), 4 \omega_{np}^2,
2(\omega_p^2+\omega_{np}^2)$. If we neglect the different coupling
between neutrons and protons $\omega_{np}$ we only obtain
two modes analogously to isovector and isoscalar ones. We see
that the coupling between neutrons and protons can lead to the
appearance of a third mode.

\begin{figure}
\centerline{\psfig{file=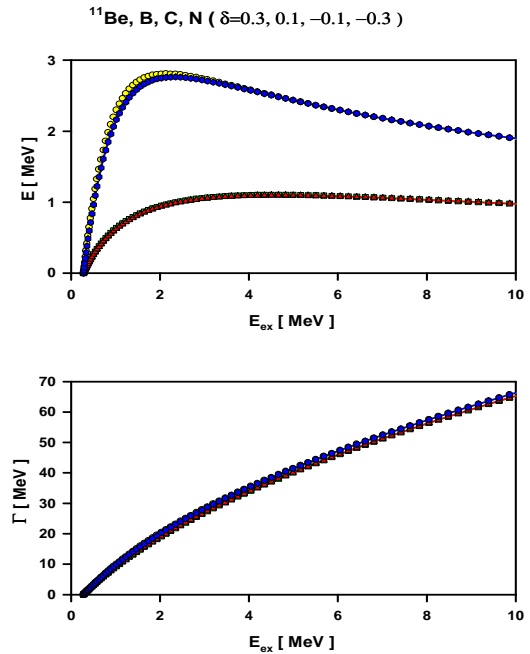,height=10cm,width=8cm,angle=0}}
 \caption{The centroid energy (above) and the damping width
 (below) of collective modes vs. excitation energy (above Fermi energy) 
for the new soft mode for different isobaric states of $^{11}Be$. The symmetric nuclei $^{11}Be$ (circles) and $^{11}N$ (rhombus) are almost equal and three times larger in the centroid energy than the more symmetric cases $^{11}B$ (triangles) and $^{11}C$ (squares). \label{3} }
\end{figure}

We conclude that due to collisional coupling there should be a
third soft collective mode observable besides the isovector and
isoscalar modes. To study further these modes may open
an exciting access to correlation effects in nuclear matter. In
figure \ref{3} we give the predictions for the isobar states of
$^{11}Be$. We find that for $^{11}B$ and $^{11}C$ the modes are by a factor three lower. Interesting to remark that the
damping width is found to be independent of the asymmetry.

Let us now compare the found new mode with the experimental evidence.
There are some hints for a soft mode in $^{11}Be$ \cite{C96}.
The authors have observed a low lying structure at around
$6$ MeV excitation energy with a damping of around $1$ MeV which has not been
reproduced yet even within refined coupled channel calculations
\cite{EBS95}. A standard explanation would give as the origin a weekly bounded single particle neutron orbital. 
The observed broad structure at $6$ MeV might
be explained possibly as the here presented coupled mode.
The centroid energy as well as damping width at least seem to
suggest this interpretation.

Again, we like to point out that the presented results are limited to a pure liquid drop model. 
We have neglected shell effects and surface effects which are important 
for realistic calculations in small systems like $^{11}Be$.

To summarize we have observed that due to correlational coupling
there can exist a new mode which appears besides isovector and
isoscalar modes in asymmetric nuclear matter due to collisions. 
We suggest that
this mode may be possible to observe as a soft collective
excitation in asymmetric systems. The transition from zero sound damping to first sound damping behavior should become possible to observe for isovector modes since they do not vanish at this transition temperature like in symmetric matter.

\acknowledgements
The authors thank for the fruitful discussions
with M. DiToro, A. Larionov and V. Baran.
The hospitality of LNS-INFN Catania where part of this work has
been
finished, is gratefully acknowledged by K. M. and R. W.
The work was supported by the Max-Planck Society Germany and the DFG under contract Nr. Ro905/13-1.


\end{document}